\documentclass[journal,twoside,web]{ieeecolor}
\usepackage{tmi}
\usepackage{cite}
\usepackage{amsmath,amssymb,amsfonts}
\usepackage{algorithmic}
\usepackage{graphicx}
\usepackage{subfigure}
\usepackage{textcomp}
\usepackage{todonotes}\def\BibTeX{{\rm B\kern-.05em{\sc i\kern-.025em b}\kern-.08em
    T\kern-.1667em\lower.7ex\hbox{E}\kern-.125emX}}
\begin{document}
\title{Joint Calibrationless Reconstruction and Segmentation of Parallel MRI}
\author{Aniket Pramanik, \IEEEmembership{Student Member, IEEE}, Xiaodong Wu, and Mathews Jacob, \IEEEmembership{Senior Member, IEEE}
\thanks{Aniket Pramanik, Xiaodong Wu and Mathews Jacob are from the
Department of Electrical and Computer Engineering at the University of
Iowa, Iowa City, IA, 52242, USA (e-mail: aniket-pramanik@uiowa.edu; xiaodong-wu@uiowa.edu; mathews-jacob@uiowa.edu). This work
is supported by NIH  grants R01EB019961 and R01AG067078.}
}

\maketitle

\begin{abstract}
The volume estimation of brain regions from MRI data is a key problem in many clinical applications, where the acquisition of data at high spatial resolution is desirable. While parallel MRI and constrained image reconstruction algorithms can accelerate the scans, image reconstruction artifacts are inevitable, especially at high acceleration factors. We introduce a novel image domain deep-learning framework for calibrationless parallel MRI reconstruction, coupled with a segmentation network to improve image quality and to reduce the vulnerability of current segmentation algorithms to image artifacts resulting from acceleration. The combination of the proposed image domain deep calibrationless approach with the segmentation algorithm offers improved image quality, while increasing the accuracy of the segmentations. The novel architecture with an encoder shared between the reconstruction and segmentation tasks is seen to reduce the need for segmented training datasets. In particular, the proposed few-shot training strategy requires only 10\% of segmented datasets to offer good performance.
     
\end{abstract}


\section{Introduction}
\label{sec:introduction}
The degeneration/atrophy of brain structures is a key biomarker that is predictive of progression in several neurodegenerative disorders, including Alzheimer's disease \cite{de2015structural,carlesimo2015atrophy,chetelat2018multimodal}. The volume measures of these brain structures are usually estimated in the clinical setting from the segmentation of MR images. High-resolution images can improve the precision of volume measures and thus enable the early detection and improve the prediction of progression \cite{pruessner2000volumetry,iglesias2015computational,lusebrink2013cortical}. Unfortunately, MRI is a slow imaging modality; the acquisition of high-resolution images often comes with a proportional increase in scan time, which is challenging for several patient sub-groups, including older adults. In addition, longer scans are vulnerable to motion artifacts \cite{zaitsev2015motion}. Calibrated \cite{pruessmann1999sense,griswold2002generalized} parallel MRI (PMRI) schemes, which use pre-estimated coil sensitivities, and calibration-free PMRI schemes \cite{shin2014calibrationless,haldar2013low,mani2017multi} have been used to accelerate the acquisition. While calibrationless approaches can eliminate motion artifacts and offer higher acceleration factors, the high computational complexity of these methods is a limitation. While current PMRI schemes use constrained models including compressed sensing \cite{lustig2007sparse,liang2009accelerating} and deep-learning priors \cite{hammernik2018learning,aggarwal2018modl}, the reconstructed images often exhibit residual aliasing and blurring at high acceleration rates. The direct use of current segmentation algorithms on these images may result in segmentation errors, which may offset the benefit of the higher spatial resolution. In particular, we note that the reconstruction and segmentation algorithms are often designed and developed independently, even though there is extensive synergy between segmentation and reconstruction tasks. A challenge that restricts the development of joint recovery-segmentation algorithms is the lack of datasets with segmentation labels. In particular, segmentation often requires extensive time and expert supervision; a semi-supervised segmentation approach that can reduce the number of labelled datasets can significantly improve the widespread adoption of such methods. 

The main focus of this paper is to introduce a deep-learning framework for joint calibrationless reconstruction and semi-supervised segmentation of PMRI data. Motivated by our previous k-space deep-learning (DL) strategy \cite{pramanik2020deep} for calibrationless PMRI, we introduce a novel image domain DL framework. The proposed framework is motivated by the CLEAR approach \cite{trzasko2012clear}, which exploits the low-rank structure of the patches obtained from coil sensitivity weighted images. To reduce the computational complexity, we unroll an iterative re-weighted least squares (IRLS) algorithm to minimize the CLEAR cost function and train it end-to-end, similar to the approach in \cite{aggarwal2018modl,hammernik2018learning,schlemper2017deep} with shared weights across iterations as in model-based deep-learning (MoDL) \cite{aggarwal2018modl}. We note that in an IRLS-CLEAR formulation, the iterations alternate between a data consistency (DC) and a projection step, which projects each set of multi-channel patches to a linear signal subspace, thus \emph{denoising} them. The IRLS algorithm estimates the linear subspaces or, equivalently, the linear annihilation relations from the data itself, which requires several iterations that contribute to the high computational complexity. We instead replace the linear projection step with a residual UNET \cite{ronneberger2015u} convolutional neural network (CNN). We hypothesize that the pre-learned non-linear CNN acts as a spatially varying annihilation filterbank for the multi-channel patches. The use of the CNN, whose parameters are pre-learned from exemplar data, translates to good performance with much fewer iterations and, hence, significantly reduced computational complexity. We note that significantly more inter- and intra-patch annihilation relations are available in the spatial domain compared to the k-space approach in \cite{pramanik2020deep}, which translates to improved performance. 

To make the segmentation scheme robust to blurring and aliasing artifacts at high acceleration rates, we propose an integrated framework for the joint segmentation and reconstruction. In particular, we attach a segmentation-dedicated decoder to the encoder of the projection CNN at the final iteration of the unrolled reconstruction network. Please see Fig. \ref{fig:prop_arch}. The shared architecture, which uses the common latent representation for both the segmentation and \emph{denoising} at all the iterations, facilitates the exploitation of the synergies between the tasks. More importantly, this approach enables a semi-supervised training strategy that significantly reduces the number of segmented datasets that are needed to train the joint reconstruction-segmentation framework. We also expect the reconstruction task to benefit from the addition of the segmentation task, which emphasizes the accuracy of the edge locations. 
In the fully supervised setting, the multi-task architecture is trained end-to-end; the loss is the weighted linear combination of mean squared reconstruction error (MSE) and multi-class cross entropy segmentation loss. We also introduce a semi-supervised setting to reduce the number of segmented datasets. In this setting, we use a weighted linear combination of reconstruction and segmentation losses for the datasets with segmentation labels. By contrast, we only use the reconstruction MSE loss to {\rm train} the network for datasets without segmentation labels. The shared encoder, which is trained in the reconstruction task on all subjects, enables us to keep the  generalization error for segmentation low, even when few labelled segmentation datasets are used. This semi-supervised approach is an attractive option, especially since there are not many large publicly available datasets with both raw k-space data and segmentation labels. 

 The proposed work has conceptual similarities to the cascade of reconstruction and segmentation networks \cite{huang2019fr} and the recent JRS \cite{sun2019joint}. The use of the same latent spaces in our approach facilitates the improved exploitation of the synergies, which offers improved performance and enables the semi-supervised approach rather than the cascade approach \cite{huang2019fr}. In addition to being restricted to the single-channel MRI setting, the JRS framework has some important distinctions from the proposed approach. The parameters of the auto-encoders at each iteration are not shared in \cite{sun2019joint}. To constrain the latent variables at each iteration, they are each fed to a segmentation network; the segmentation errors from each iteration are combined linearly to define the segmentation loss. Unlike \cite{sun2019joint}, we use the same reconstruction network at each iteration and only connect the segmentation network to the latent variables in the final iteration. The sharing of the parameters in our setting ensures that the segmentation task will regularize the projection networks at all iterations, thus facilitating the semi-supervised training strategy. 
\section{Background}
\subsection{Forward Model}
Parallel MRI acquisition of $\gamma$ can be modelled as
\begin{equation}
b_{i} = \mathcal U(\mathcal F(\underbrace{s_i  ~  \gamma}_{\gamma_i})) +  \eta_i, \hspace{2pt} i = 1,\ldots,N
\end{equation}
where $N$ is the number of receiver coils, $s_i$ is the sensitivity of the $i^{\rm th}$ coil, $b_i$ represents the undersampled measurements corresponding to the $i^{\rm th}$ coil, and $\eta_i$ is zero mean additive white Gaussian noise. Here, $\mathcal F$ is a 2D fast Fourier transform (FFT) operator that transforms the spatial domain signal $\gamma_i$ to its k-space, which is sampled at the locations defined by the undersampling operator $\mathcal U$. 

The model can be compactly represented as $\mathbf b = \mathcal A(\boldsymbol \gamma) + \mathbf n$, where $\boldsymbol{\gamma}$ is a 3D volume obtained by stacking the coil images ${\gamma_1}, .. ,{\gamma}_N 
$ in the spatial domain, $\mathbf b$ is a concatenation of corresponding noisy undersampled Fourier measurements $b_1, .. , b_N$ across channels and $\mathbf n$ is a concatenation of corresponding noise $\eta_1, .. , \eta_N$ for the $N$ channels. Here, $\widehat{\boldsymbol \gamma} = \mathcal F(\boldsymbol \gamma)$ is multi-channel data in the Fourier domain and, similarly, the Fourier domain image $\widehat{\gamma_i} = \mathcal F(\gamma_i)$ for $i^{\rm th}$ channel.   
\subsection{Calibrationless PMRI Recovery}
Parallel MRI recovery methods can be broadly classified into calibrated and calibrationless approaches. Calibrated approaches such as SENSE \cite{pruessmann1999sense}, GRAPPA \cite{griswold2002generalized}, and ESPiRIT \cite{uecker2014espirit} have a separate coil sensitivity estimation step, either through additional calibration scans or from self-calibration data. Recently, MoDL methods have been used to further improve the performance of the above calibrated PMRI schemes \cite{hammernik2018learning,aggarwal2018modl}. These methods pose the recovery as an optimization problem, which minimizes the sum of a DC term and a deep-learned CNN prior. An alternating minimization scheme to solve the above problem is unrolled for a finite number of iterations and trained in an end-to-end fashion \cite{aggarwal2018modl}. These approaches offer improved performance over classical calibrated PMRI schemes. A challenge with the above calibrated schemes is their sensitivity to the estimated sensitivity maps; any imperfections in the estimated sensitivities introduces error in reconstructions. 

Calibrationless recovery methods were introduced to minimize the above model mismatch. These approaches, often termed as structured low-rank (SLR) methods, exploit the low-rank property of multi-channel k-space patches to recover the multi-channel images from undersampled data \cite{shin2014calibrationless,haldar2013low,mani2017multi}. Another SLR method, CLEAR \cite{trzasko2012clear}, was introduced to recover PMRI by utilizing relations in the spatial domain instead of k-space. It works on a locally low-rank assumption where a matrix containing patches extracted from images across the coils is low-rank. The spatial relations are obtained based on the smoothness of coil sensitivities. While these methods are powerful, the main challenge is their high computational complexity; it often takes several minutes to recover slices of a subject. Recently, DL methods were introduced for faster calibrationless recovery. For example, K-UNET \cite{han2019k} is a multi-channel UNET used to map undersampled coil images directly to their corresponding fully sampled ones. A similar approach, called Deep-SLR, is used in the model-based setting and offers improved performance over the data-driven method \cite{han2019k}.

\subsection{End-to-end Multi-task Training Approaches}
Most of the MRI segmentation algorithms are designed for fully sampled data. When the images are acquired with high acceleration factors, the recovered images often exhibit significant artifacts, including aliasing artifacts and blurring. The direct application of the above segmentation algorithms to these images may result in segmentation errors. Recently, some researchers have looked into minimizing these errors by coupling image denoising or reconstruction tasks with segmentation, thus improving both tasks  \cite{liu2018image}. \cite{huang2019fr} considers the cascade of reconstruction and segmentation networks, which are trained end-to-end, while \cite{sun2019joint} introduces an architecture as shown in Fig. \ref{fig:arch_comp}.(b) in the single-channel MRI setting.
Similar multi-task learning strategies are also used in the MRI setting \cite{oksuz2020deep}; these studies show that when the tasks are complementary, the joint learning of them using a single network can facilitate the exploitation of the synergies.  

\subsection{Semi-supervised learning}
The scarcity of labelled segmentation datasets has prompted some researchers to introduce semi-supervised learning strategies. A recent work proposes a few-shot learning approach for segmentation, which couples the segmentation task with few labelled datasets, with an image denoising task \cite{feyjie2020semi}. The segmentation and denoising tasks share the encoder; the training of the shared encoder-denoising decoder on all the datasets allows them to keep the generalization error of the segmentation task low, even though few segmentation labels were used for training. Another work proposes few-shot learning for an image classification task, which is coupled with a self-supervised auxiliary task \cite{gidaris2019boosting}.   
\section{Proposed Framework}
We will first introduce the image domain deep-SLR (IDSLR) scheme, which offers improved recovery over the k-space deep-SLR in \cite{pramanik2020deep}. We will then combine the reconstruction scheme with a semi-supervised segmentation scheme to reduce the sensitivity to alias artifacts and blurring associated with high acceleration factors.
\subsection{Image-Domain Deep-SLR}
We describe the CLEAR approach and an IRLS algorithm to solve it. Subsequently, we will propose an alternating minimization strategy, which culminates into an IDSLR network. 
\subsubsection{Spatial Domain Annihilation Relations in CLEAR}
Consider a patch extraction operator $P_{\mathbf c}$, which extracts $M \times M$ patches centered at $\mathbf c$ from the multi-channel data $\gamma_i(\mathbf r) = s_i(\mathbf r) \gamma(\mathbf r)$, where $\mathbf r$ denotes spatial coordinates. The coil sensitivity functions $s_i(\mathbf r)$ are smooth functions and hence can be safely approximated as $s_i(\mathbf r)\approx s_i(\mathbf c); \forall \mathbf r$ within the $M\times M$ patch neighborhood. The CLEAR approach relies on this approximation to show that the patch matrices 
\begin{eqnarray}\label{patches}
\boldsymbol\Gamma_{\mathbf c} &=& \begin{bmatrix}
P_{\mathbf c}(\gamma_1), & \ldots & ,P_{N}(\gamma_N)
\end{bmatrix}\\
&\approx& P_{\mathbf c}(\gamma)\begin{bmatrix}s_1(\mathbf c),&s_2(\mathbf c),&\ldots&,s_N(\mathbf c)
\end{bmatrix}
\end{eqnarray}
can be approximated as rank 1 matrices \cite{trzasko2012clear}. The dimensions of $\boldsymbol\Gamma_{\mathbf c}$ are $M^2 \times N$ with $N < M^2$. We note that when the size of the patches increases, the approximation $s_i(\mathbf r)\approx s_i(\mathbf c)$ breaks down. However, the coil sensitivities still may be well approximated as the sum of few principal components within the patches; the matrices $\boldsymbol\Gamma_{\mathbf c}$ will be low-rank even in those cases. CLEAR solves for 3-D multi-channel volume $\boldsymbol \gamma$ by minimizing the nuclear norm optimization problem \cite{trzasko2012clear}:
\begin{equation}
\label{clear}
\boldsymbol \gamma = \arg \min_{\boldsymbol\gamma} \|\mathcal A(\boldsymbol\gamma) - \mathbf b\|_2^2 + \lambda \sum_c \|\boldsymbol \Gamma_{\mathbf c}\|_{\ast}
\end{equation}

If the rank of $\boldsymbol\Gamma_{\mathbf c}$ is $r < N$, there are $N-r$ null space vectors $\mathbf u_{\mathbf c,j}$ and $\mathbf v_{\mathbf c,j}$ such that 
\begin{equation}\label{annihilation}
\mathbf u_{\mathbf c,j}~ \boldsymbol\Gamma_{\mathbf c}=\boldsymbol\Gamma_{\mathbf c}~\mathbf v_{\mathbf c,j}=0; ~j=0,..,N-r
\end{equation}
If we consider the vertical concatenation of the patches denoted by the vector $\mathbf p_{\mathbf c}$, we can express the relation in \eqref{annihilation} compactly as 
\begin{equation}\label{mtxform}
\underbrace{\begin{bmatrix}
\mathbf u_{\mathbf c,j}&\ldots&0\\
\vdots&\vdots&\vdots\\
0&0&\mathbf u_{\mathbf c,j}\\
\hline\\
\mathbf v_{\mathbf c,j}(1)\mathbf e_0&\ldots& \mathbf v_{\mathbf c,j}(M^2)\mathbf e_0\\
\vdots&\vdots&\vdots\\
\mathbf v_{\mathbf c,j}(1)\mathbf e_{M^2}&\ldots& \mathbf v_{\mathbf c,j}(M^2)\mathbf e_{M^2}\\
\end{bmatrix}}_{\mathbf Q_{\mathbf c}}
\underbrace{\begin{bmatrix}P_{\mathbf c}(\gamma_1)\\
\vdots\\
P_{\mathbf c}(\gamma_N)\\
\end{bmatrix}}_{\mathbf p_{\mathbf c}} = \boldsymbol 0
\end{equation}
Here, $\mathbf e_i$ denotes the canonical basis vectors of the patches. The top set of $N$ relations indicate intra-channel annihilation relations that work independently on the patch from each channel, while the bottom set of $M^2$ relations correspond to inter-channel annihilation relations. Unlike the setting in k-space based SLR methods, these annihilation relations are spatially varying. These relations can be can be viewed as the convolution of the 3D multi-channel volume $\boldsymbol \gamma(\mathbf r)$, whose third dimension is the channels $\gamma_1(\mathbf r),..,\gamma_N(\mathbf r)$, with flipped filters of size $M\times M\times N$ that are spatially varying, evaluated at $\mathbf c$:
\begin{equation}
\label{multichannelconv}
\boldsymbol\gamma(\mathbf r)*\mathbf q_{\mathbf c,j,i}(\mathbf r)\Big|_{\mathbf c}=0; \hspace{2pt} i=0,\ldots,N+M^2
\end{equation}
are multi-channel filters corresponding to the rows of $\mathbf Q_{\mathbf c}$ in \eqref{mtxform}.

\subsubsection{Iterative Re-weighted Least Squares (IRLS) Formulation}
An IRLS approach to \eqref{clear} results in an alternating minimization algorithm that alternates between the solution of the images:
\begin{equation}
\label{denoising}
\arg \min_{\boldsymbol \gamma} \|\mathcal A(\boldsymbol \gamma) - \mathbf b\|_2^2
+ \lambda \sum_{\mathbf c,i,j}\|(\mathbf q_{\mathbf c,j,i} \ast \boldsymbol \gamma)|_{\mathbf c}\|_F^2
\end{equation}
and deriving the spatially varying multichannel filters $\mathbf q_{\mathbf c,j,i}$ of size $M\times M\times N$ in \eqref{multichannelconv}. These filters are estimated from the signal patch matrices $\boldsymbol \Gamma_{\mathbf c}$. The regularization term in \eqref{denoising} minimizes the projection of the signal energy to the null-space.   

\subsubsection{Image domain Deep-SLR formulation}
\begin{figure}[t!]
	\centering
	\includegraphics[width=0.48\textwidth,keepaspectratio=true,trim={0.5cm 6cm 0.5cm 6cm},clip]{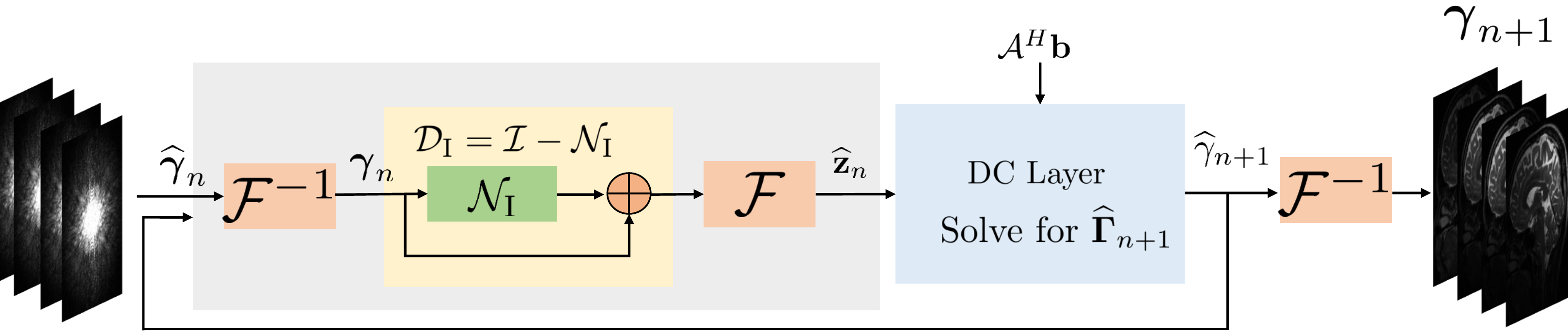}
	\caption{Proposed IDSLR network: It consists of residual
CNN $\mathcal D_{\rm I}$ for spatial domain learning. $\mathcal D_{\rm I}$ exploits the image domain annihilation relations among patches to denoise the multi-channel image $\boldsymbol \Gamma$. The network is unrolled for $K$ iterations and trained end-to-end. Output of $\mathcal D_{\rm I}$ at the $n^{\rm th}$ iteration is denoted by $\mathbf Z_n$ according to \eqref{den}. The network parameters are shared across iterations similar to Deep-SLR \cite{pramanik2020deep} framework.}

	\label{fig:prop_idslr}
\end{figure}

Similar to \cite{pramanik2020deep}, an alternating minimization scheme is derived to obtain a recursive network architecture that iterates between a data consistency enforcement (DC) and a denoising step. Motivated by our k-space DL framework \cite{pramanik2020deep}, we replace the linear null-space projection step with a non-linear spatial domain CNN that can learn a set of non-linear spatially varying annihilating filters from the training data and generalize well over unseen test data. The proposed formulation is
\begin{equation}
\label{idslr}
\boldsymbol \gamma = \arg \min_{\boldsymbol \gamma} \|\mathcal A(\boldsymbol \gamma) - \mathbf b\|_2^2 + \lambda \|\underbrace{(\mathcal I - \mathcal D_{\rm I}}_{\mathcal N_{\rm I}})(\boldsymbol \gamma)\|_2^2
\end{equation}  
where $\mathcal N_{\rm I} = \mathcal I - \mathcal D_{\rm I}$ denotes a multi-channel spatial domain CNN and $\mathcal I$ is the identity operator. Note the similarity of \eqref{idslr} to \eqref{denoising}; we expect the non-linear spatial domain CNN $\mathcal N_{\rm I}$ to realize the spatially varying convolutions $\mathbf q_{\mathbf c,j,i}*\boldsymbol \gamma$. The solution to \eqref{idslr} alternates between the residual CNN $\mathcal D_{\rm I}$ and the DC block as shown in Fig. \ref{fig:prop_idslr}. The corresponding equations are
\begin{eqnarray}
\label{den}
\mathbf z_n &=& \mathcal D_{\rm I}(\boldsymbol \gamma_n) \\
\label{dc}
\widehat{\boldsymbol \gamma}_{n+1} &=& (\mathcal A^H \mathcal A + \lambda \mathcal I)^{-1}(\mathcal A^H\mathbf b + \lambda \widehat{\mathbf z}_n) 
\end{eqnarray}
The recursive network structure is shown in Fig. \ref{fig:prop_idslr}, which alternates between the residual CNN $\mathcal D_{\rm I}$ and the DC block. We propose to learn the CNN parameters by unrolling the algorithm for a finite number of iterations $K$ and training it end-to-end using exemplar datasets. Similar to \cite{pramanik2020deep}, the CNN weights are shared across iterations. An MSE loss is computed between the predicted image $\boldsymbol \gamma$ and the gold standard image $\boldsymbol \gamma^{gs}$ obtained from fully sampled measurements. It can be written as $\mathcal L_{\rm rec} = \|\boldsymbol \gamma_s - \boldsymbol \gamma^{gs}_s\|_2^2$.
Here, $\boldsymbol\gamma_s$ is the sum-of-squares image obtained from the multi-channel reconstructed image $\boldsymbol\gamma$, while $\boldsymbol\gamma^{gs}_s$ is the reference (gold standard) sum-of-squares image. 

The proposed algorithm does not require any calibration data for estimation of coil sensitivities or linear filters $\mathbf Q_{\mathbf c}$. Computationally, this approach is orders of magnitude faster than CLEAR since the unrolled algorithm requires fewer iterations to converge. In addition, it eliminates the need for performing singular value decomposition (SVD) of $\boldsymbol \Gamma_{\mathbf c}$ every iteration, which increases the computational complexity of CLEAR.

Apart from the multi-channel annihilation relations \eqref{annihilation} resulting from the multi-channel acquisition, MR images also satisfy annihilation properties resulting from the redundancy of patches. For instance, smooth regions can be annihilated by derivative filters, while wavelet filters are effective in annihilating texture regions. The proposed image domain multi-channel annihilation CNN framework can exploit these annihilation relations as well, which is a distinction from the k-space approach in \cite{pramanik2020deep}; the larger number of annihilation relations can translate to improved performance. Note that the traditional CLEAR approach only exploits the multi-channel relations in \eqref{annihilation}.

\subsection{Joint Reconstruction-Segmentation Framework}
MRI reconstruction and segmentation are often studied as two different tasks in the MR image analysis pipeline. In the case of a direct cascade of tasks, any aliasing or blurring artifact in reconstructions due to highly accelerated acquisition gets propagated to the segmentation task, leading to degradation of segmentation quality \cite{sun2019joint}. In this section, we propose a multi-task framework for joint reconstruction and segmentation to obtain segmentations that are robust to undersampling artifacts. We also expect the segmentation task to aid the reconstructions. 

\subsubsection{Shared Encoder Architecture}
The proposed multi-task DL framework is shown in Fig. \ref{fig:prop_arch}. The unrolled version of the IDSLR network in Fig. \ref{fig:prop_idslr} with shared weights across iterations is used for reconstruction, together with an additional segmentation network. For IDSLR, the CNN $\mathcal N_{\rm I}$ is a UNET with encoder $\mathcal N_{\rm IE}$ and decoder $\mathcal N_{\rm ID}$. A segmentation task dedicated decoder $\mathcal S_{\rm ID}$ is connected to the output of the $\mathcal N_{\rm IE}$ from the final iteration of IDSLR. Thus, the encoder for segmentation UNET is $\mathcal N_{\rm IE}$, whose parameters are shared with the reconstruction task. We expect that the shared encoder will encourage the learning of common latent variables for both the segmentation and reconstruction tasks, thus exploiting the synergy between them. In particular, we hypothesize that the spatial domain features extracted by the encoder $\mathcal N_{\rm IE}$ from multiple channels would become sharper due to the auxiliary segmentation loss, resulting in improved representation of edges and finer details. The shared encoder architecture also facilitates the learning of the segmentation task from a few labelled datasets as explained below. 

\subsubsection{Loss Function}
A weighted linear combination of reconstruction and segmentation losses is used to {\rm train} the multi-task network end-to-end.  Let $\boldsymbol\theta$, $\boldsymbol\phi$ and $\boldsymbol\psi$ be the weights of $\mathcal N_{\rm IE}$, $\mathcal N_{\rm ID}$, and $\mathcal S_{\rm ID}$, respectively. 
We used the MSE loss,
\begin{equation}
\label{recloss}
 \mathcal L_{\rm rec}(\boldsymbol\gamma^{gs}_s,\boldsymbol\gamma_s;\boldsymbol\theta,\boldsymbol\phi) = \|\boldsymbol\gamma^{gs}_s - \boldsymbol\gamma_s(\boldsymbol\theta,\boldsymbol\phi)\|^2_2 
\end{equation}
as the reconstruction loss. The output of the reconstruction network is only dependent on $\boldsymbol\theta$ and $\boldsymbol\phi$. The segmentation decoder outputs probability maps of the tissue classes denoted by $\boldsymbol p(\boldsymbol\theta,\boldsymbol\phi,\boldsymbol \psi)$; note that the probability maps are dependent on all the network parameters. We compare $\boldsymbol p$ against the reference segmentation $\boldsymbol z(\mathbf r)$ using the pixel-level multi-label cross-entropy loss
\begin{equation}
\label{segloss}
 \mathcal L_{\rm seg}(\boldsymbol p,\boldsymbol z; \boldsymbol\theta,\boldsymbol\phi,\boldsymbol \psi) = -\sum_{\mathbf r} \boldsymbol z(\mathbf r)  \log(\boldsymbol p(\mathbf r;\boldsymbol\theta,\boldsymbol\phi,\boldsymbol\psi)). 
\end{equation}
Here, $\boldsymbol p_i(\mathbf r;\boldsymbol\theta,\boldsymbol\phi,\boldsymbol \psi)$ is the predicted probabilities at the pixel $\mathbf r$ corresponding to the $i^{\rm th}$ dataset, while $\boldsymbol z_i(\mathbf r)$ is the reference labels obtained from the reference image $\gamma_{s,i}^{gs}(\mathbf r)$. The total loss is given as
\begin{eqnarray}
\label{loss}
\nonumber \mathcal L(\boldsymbol\theta, \boldsymbol\phi,\boldsymbol\psi) = \sum_{i=1}^{N_t} ~(1-\alpha) ~\mathcal L_{\rm rec}(\boldsymbol \gamma_{s,i}^{gs},\boldsymbol \gamma_{s,i})\\ + \alpha ~\mathcal L_{\rm seg}(\boldsymbol p_i,\boldsymbol z_i),
\end{eqnarray} 
where $0\leq \alpha < 1$ is the weight term that controls the strength of each term and $N_t$ is the total number of training datasets. Here, $\mathcal L_{\rm rec}$ and $\mathcal L_{\rm seg}$ are specified by \eqref{recloss} and \eqref{segloss}, respectively. 

\subsection{Few-Shot Learning for Semantic Segmentation}
\begin{figure*}[t!]
	\centering
	\includegraphics[width=\textwidth,keepaspectratio=true,trim={0cm 4cm 0cm 4.5cm},clip]{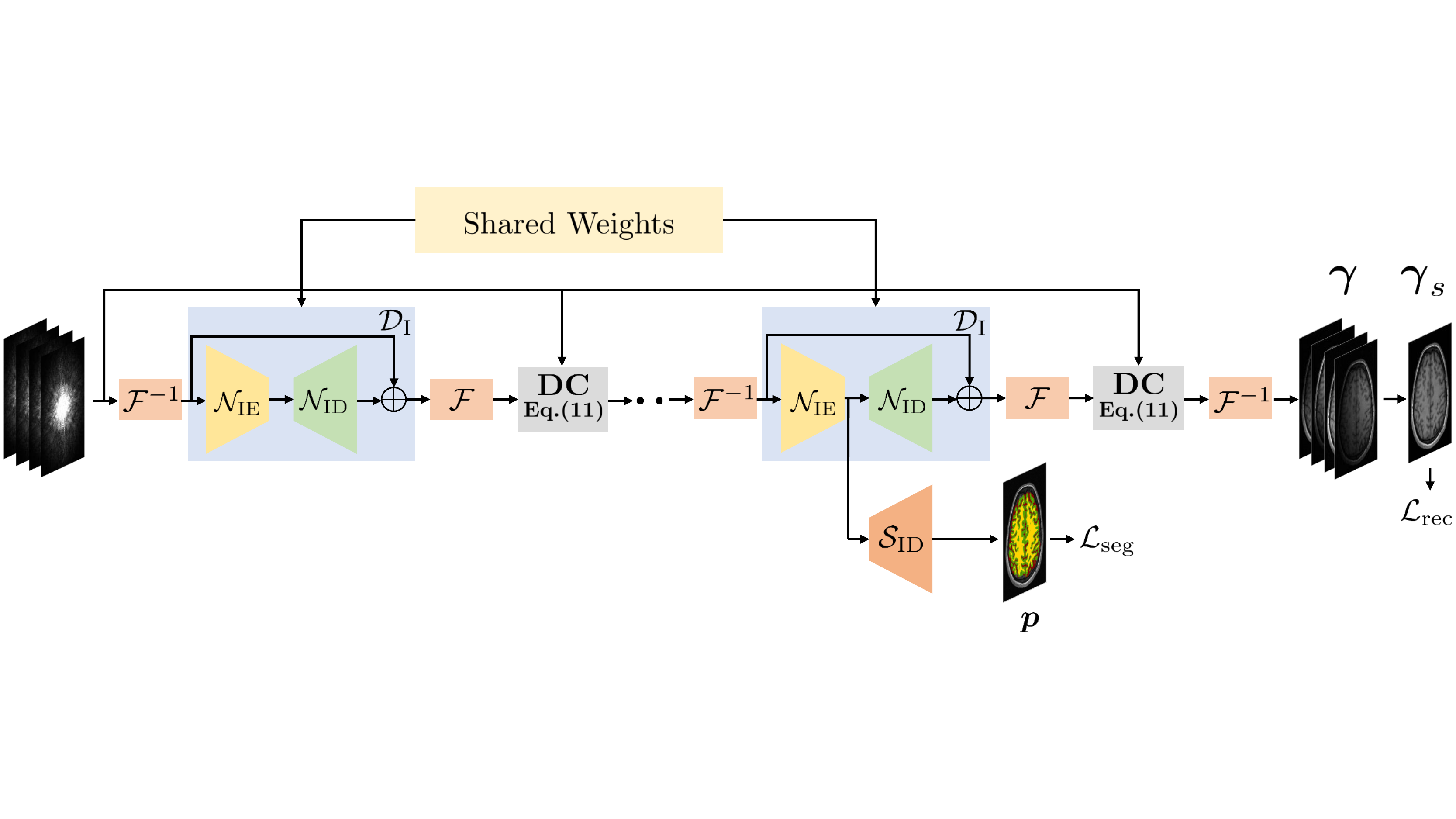}
	\caption{Proposed reconstruction-segmentation architecture. It consists of a $K$-iteration IDSLR network with the residual CNN $\mathcal D_{\rm I}$. We choose a UNET with skipped connections between encoder $\mathcal N_{\rm IE}$ and decoder $\mathcal N_{\rm ID}$. We connect the output of $\mathcal N_{\rm IE}$ from the $K^{\rm th}$ iteration to the segmentation decoder $\mathcal S_{\rm ID}$. There are skipped connections between $\mathcal N_{\rm IE}$ and $\mathcal S_{\rm ID}$, thus forming a UNET architecture for segmentation. The CNN weights are shared across iterations. It is trained end-to-end with a linear combination of reconstruction and segmentation losses.}
	\label{fig:prop_arch}
\end{figure*}

Current DL models for medical image segmentation are often trained with large amounts of training data. However, obtaining accurate segmentation labels is often a challenging task, requiring high computation time and expert supervision. On the other hand, using a few datasets to train current networks is often associated with the risk of over-fitting. In order to address the limited availability of labelled data, we propose a semi-supervised learning approach capitalizing on our shared network shown in Fig. \ref{fig:prop_arch}. In particular, we utilize the reconstruction as an auxiliary task for spatial feature learning. We are motivated by \cite{feyjie2020semi}, where an image denoising block is added as an auxiliary task to keep the generalization error of the segmentation task low.       

The proposed multi-task network is trained end-to-end in a supervised fashion with few labelled subjects ($10\%$ of training data) for segmentation. We use the combined loss function specified by \eqref{loss} for datasets with segmentation labels, while we use the reconstruction loss specified by \eqref{recloss} for datasets without segmentation labels. The shared encoder facilitates the learning of latent features that are useful for both tasks. The latent variables derived by the shared encoder are learned from all the datasets. Because the segmentation task relies on these latent variables, a few labelled segmentation datasets are sufficient to obtain good generalization performance.


\section{Implementation Details}
\subsection{Datasets}
The publicly available Calgary Campinas (CCP) Dataset \cite{souza2018open} is used for all the experiments. It consists of raw k-space for T1-weighted brain MRI datasets from 117 subjects acquired using a 12-channel coil with a gradient-echo sequence. The data was collected on a General Electric (GE) 3T scanner (Discovery MR750). The acquisition parameters are either TR (repetition time) = 6.3 ms, TE (echo time) = 2.6 ms, TI (inversion time) = 650 ms or TR = 7.4 ms, TE = 3.1 ms, TI = 400 ms. The acquisition matrix size is 256 x 208 x 170 or 256 x 208 x 180 at a slice thickness of 1 mm with 170/180 being the slice-encoding direction. We chose a subset of subjects where fully sampled reconstruction references are available for training and performance analysis. The dataset is split into 40 subjects for training, 7 for validation and 15 for testing. CCP does not provide reference segmentation corresponding to the raw k-space data for any subject. We generated reference segmentation using FMRIB's Automated Segmentation Tool (FAST) software \cite{zhang2001segmentation}. It uses a hidden Markov random field model and solves an associated Expectation-Maximization algorithm to classify the image pixels into CSF, GM and WM. This segmentation task was chosen to demonstrate the technical feasibility of the proposed framework mainly because of the availability of well-established software that is directly applicable to the above dataset with raw multi-channel k-space data. We note that, while several segmentation datasets with more challenging tasks are publicly available, most of them rely on post-processed DICOM images that are not appropriate for our setting. 
\subsection{Quality Evaluation Metric}
We quantitatively evaluate reconstruction quality using the signal-to-noise ratio (SNR) and structural similarity \cite{wang2004image} (SSIM) metrics. The SNR of an image is computed as $
\mathbf{SNR (dB)} = 20 \cdot \log_{10} \left(\frac{\|\mathbf x_{\rm org}\|_2}{\|\mathbf x_{\rm org}-\mathbf x_{\rm rec}\|_2}\right)$, where $\mathbf x_{\rm rec}$ is the predicted image and $\mathbf x_{\rm org}$ denotes the corresponding ground truth image. The segmentation performance is evaluated using Dice coefficients \cite{jadon2020survey}.
\subsection{Architecture of the CNNs and Training Details}
The IDSLR network in Fig. \ref{fig:prop_arch} takes a multi-channel raw k-space $\boldsymbol{\widehat{\gamma}}$ as input. It goes through an inverse fast Fourier transform (IFFT) operation before entering the $\mathcal D_{\rm I}$ residual block. The $\mathcal D_{\rm I}$ block consists of a residual UNET $\mathcal N_{\rm I}$, which  consists of 20 layers with four max pooling and unpooling operations. The number of filters per layer of $\mathcal N_{\rm I}$ grows from 36 to a maximum of 512. A convolution layer with filter size 3 x 3 is followed by ReLU non-linearity. Similar to the standard UNET, there are skipped connections between the encoder $\mathcal N_{\rm IE}$ and decoder $\mathcal N_{\rm ID}$. The segmentation decoder $\mathcal S_{\rm ID}$ is attached to $\mathcal N_{\rm IE}$ from the final iteration of the unrolled IDSLR network. There are also skipped connections to $\mathcal S_{\rm ID}$ from $\mathcal N_{\rm IE}$, similar to the standard UNET architecture.      

\begin{figure}[h!]
\subfigure[Cascade architecture]{\includegraphics[width=0.48\textwidth,keepaspectratio=true,trim={0.1cm 3.8cm 0.1cm 4cm},clip]{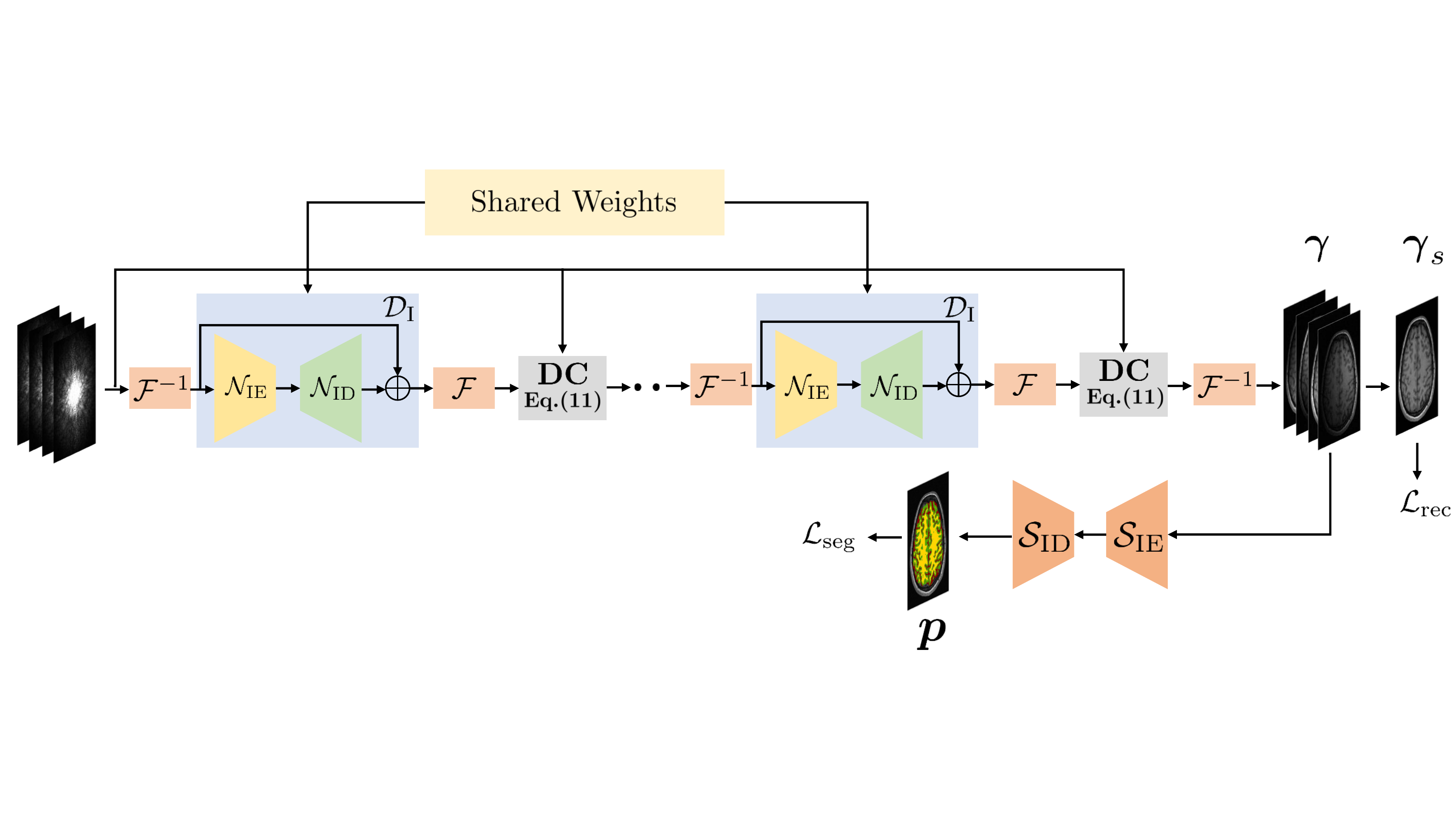}}
\subfigure[JRS architecture]{\includegraphics[width=0.48\textwidth,keepaspectratio=true,trim={0.1cm 3.9cm 0.1cm 3.9cm},clip]{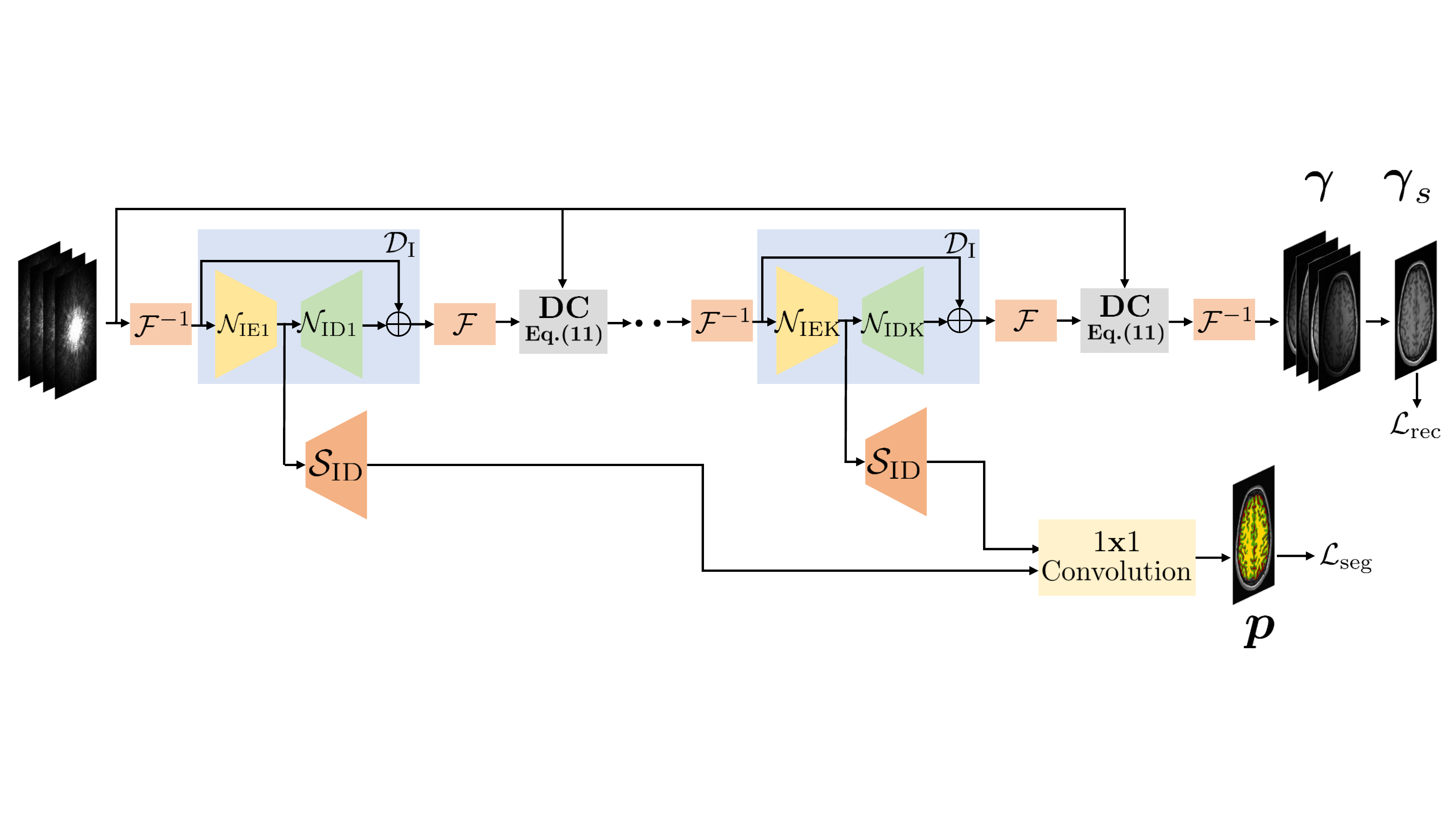}}
\caption{Architecture of joint reconstruction-segmentation frameworks that the proposed algorithm is compared against: (a) is a cascade of segmentation UNET to the proposed IDSLR framework; the joint network is trained end-to-end. A key difference with the proposed method is the segmentation encoder $\mathcal S_{\rm IE}$ for segmentation, which is unshared and results in the network having more parameters; (b) is an illustration of JRS \cite{sun2019joint} framework where weights are unshared across the $K$ iterations of reconstruction network. The reconstruction encoders and decoders are denoted by $\mathcal N_{\rm IE1}, .. , \mathcal N_{\rm IEK}$ and $\mathcal N_{\rm ID1}, .. , \mathcal N_{\rm IDK}$ respectively since the parameters are unshared. Here $\mathcal D_{\rm I}$ is a residual autoencoder instead of a UNET as in IDSLR. The segmentation weights are shared among the decoders $\mathcal S_{\rm ID}$.}
\label{fig:arch_comp}
\end{figure}     

The IDSLR network is unrolled and trained end-to-end for $K = 5$ iterations. We also fix $K = 5$ for all other iterative DL models of the k-space deep-SLR (KDSLR), JRS, and proposed and cascade architectures to make the comparisons fair. The weighting parameter $\alpha$ is chosen empirically for the proposed methods. For six-fold and eight-fold accelerations, we chose $\alpha = 10^{-4}$ and $\alpha = 10^{-6}$, respectively. The CNN weights are Xavier initialized and trained for 1000 epochs with the Adam optimizer at a learning rate of $10^{-4}$. The training data from one subject at a time is used in every epoch. All the DL methods are implemented using the Pytorch library.       

\subsection{State-of-the-art Methods for Comparison}

\subsubsection{Calibrationless PMRI methods}
The spatial domain deep-SLR (IDSLR) is compared against KDSLR \cite{pramanik2020deep}, k-space UNET \cite{han2019k} and KIKI-Net \cite{eo2018kiki} to demonstrate reconstruction performance. The K-UNET approach is a direct inversion scheme that uses a multi-channel UNET in k-space without any DC step. KIKI-Net relies on a cascade of k-space and spatial domain CNNs with a DC step embedded between them. Since the original KIKI-Net framework was designed for single-channel MRI, we implemented a multi-channel version by replacing the single-channel CNNs with multi-channel ones, together with the DC block in \eqref{dc}. KDSLR is a deep-SLR approach exploiting annihilation relations in k-space. 

\subsubsection{Combination of image recovery and segmentation}
We cascade a UNET-based segmentation network SEG, which is pre-trained on fully sampled images, with the above-mentioned recovery methods to study the effect of reconstruction on segmentation quality. KDSLR+SEG, IDSLR+SEG, K-UNET+SEG and KIKI-Net+SEG thus denote the direct cascade of the segmentation network SEG to these recovery methods without any end-to-end training.

\begin{figure*}[h!]
	\centering
	\includegraphics[width=\textwidth,keepaspectratio=true,trim={1.8cm 8.2cm 2.3cm 8.6cm},clip]{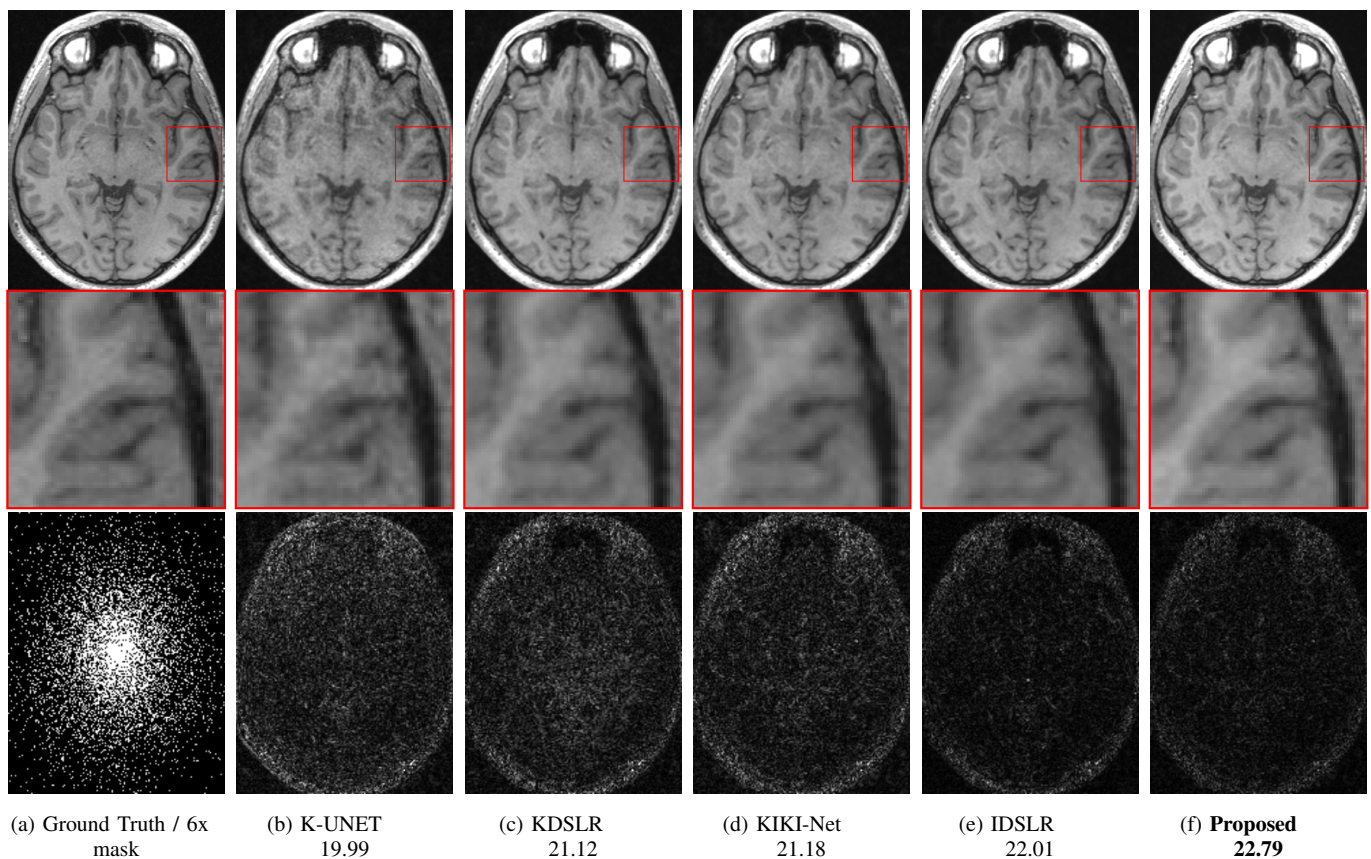}
	\caption{Reconstruction results of 6x accelerated multi-channel brain data. SNR (dB) values are reported for the particular slice in each case. The data was undersampled using a Cartesian 2D non-uniform variable-density mask. The top row shows reconstructions (magnitude images), while the bottom row shows the corresponding error images. The proposed end-to-end calibrationless scheme, which combines the segmentation and reconstruction tasks (last column), is seen to offer improved reconstructions over networks with calibrationless reconstruction alone. The proposed IDSLR scheme without the segmentation module is also seen to offer improved performance compared to classical reconstruction algorithms.}
	\label{fig:brain_rec}
\end{figure*}

\begin{figure*}[t!]
	\centering
	\includegraphics[width=\textwidth,keepaspectratio=true,trim={1.8cm 9.8cm 0.25cm 10.5cm},clip]{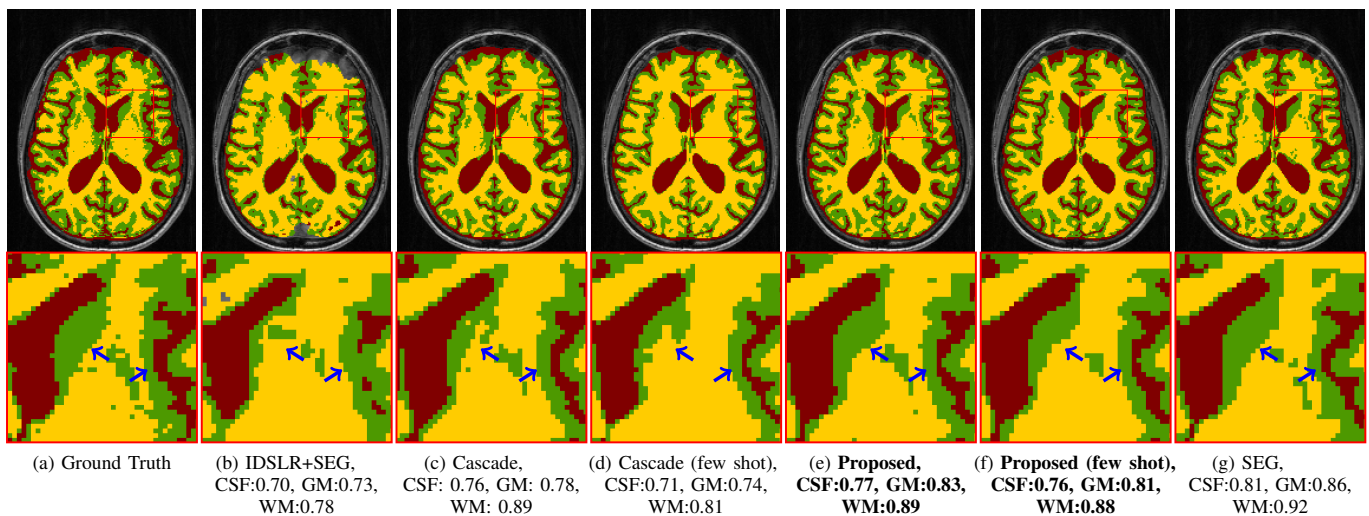}
	\caption{Comparison of segmentation performance on 8x accelerated multi-channel brain data. Dice coefficients are reported for the particular slice in each case. The segmentation algorithm pre-trained and tested on fully sampled data is denoted by SEG (last column). The second column denotes the IDSLR algorithm, cascaded with the pre-trained segmentation algorithm (SEG); the cascade is denoted as IDSLR+SEG. The remaining rows (c)-(f) denote end-to-end training strategies. (c) and (d) correspond to the cascaded architecture in Fig. \ref{fig:arch_comp}, while (e) and (f) use the architecture in Fig. \ref{fig:prop_arch}. (d) and (e) correspond to the cascade and proposed architectures trained in the few-shot mode.}
	\label{fig:brain_seg}
\end{figure*}

We also consider the joint end-to-end training strategies with the cascade model and the JRS scheme \cite{sun2019joint}, shown in Fig. \ref{fig:arch_comp}. The architecture in Fig. \ref{fig:arch_comp}.(a) is the cascade of a segmentation UNET to the IDSLR network, similar to \cite{huang2019fr}, which is trained end-to-end with the loss function specified by \eqref{loss}. For fair comparison, we match the number of learnable parameters of this approach with that of the proposed scheme. The JRS scheme \cite{sun2019joint} was originally introduced for single-channel MRI. We extended the above framework to the multi-channel setting by replacing the single-channel autoencoders with multi-channel ones and a DC block in \eqref{dc}. The JRS architecture is shown in Fig. \ref{fig:arch_comp}.(b). The key distinction of JRS with our setting is that the parameters of the autoencoders at each iteration are not shared. The latent variables at each iteration are fed to a segmentation network, and the combination of the segmentation losses at each iteration is used as the segmentation loss. Because we use the same network at each iteration, we only need to connect the segmentation network to the final layer; the sharing of the parameters will ensure that the segmentation task will regularize the projection networks at all iterations. 

\section{Experiments and Results}
We now compare the proposed framework against state-of-the-art parallel MRI as well as joint reconstruction-segmentation algorithms on brain MRI data using six- and eight-fold retrospective undersampling.  

\subsection{Calibration-free PMRI}

We report the results of the comparison of the proposed spatial domain deep-SLR scheme (IDSLR) against other calibrationless methods in Table \ref{tab:rec_comp} and Fig. \ref{fig:brain_rec}, respectively. All the schemes were tested on 12-channel brain images obtained from 15 subjects. Comparisons have been made for both six- and eight-fold undersampling. We observe that the KDSLR scheme, which uses interleaved data consistency steps along with the k-space annihilation relations used in K-UNET, can offer improved performance over K-UNET. The table shows that KIKI-Net offers a marginally improved performance over KDSLR, while IDSLR offers more than 1 dB improvement over all of the other methods. As shown in Fig. \ref{fig:brain_rec}, IDSLR is able to minimize alias artifacts while offering sharper edges. The images denoted by the proposed scheme are a combination of the proposed IDSLR approach with a segmentation network as shown in Fig. \ref{fig:prop_arch} and were jointly trained end-to-end. This approach offers an additional 0.75 dB improvement in the SNR, which can also be visualized from the zoomed-in images in Fig. \ref{fig:brain_rec}.  
\subsection{Segmentation Quality Comparison}
We compare the segmentation performance of the cascade of a segmentation network with the above networks with our proposed end-to-end multi-task learning scheme in Table \ref{tab:comp_seg} and Fig. \ref{fig:brain_seg}, respectively. The segmentation results are reported in terms of Dice coefficients averaged over 15 subjects. The SEG method consists of a UNET for segmentation, trained on fully sampled images. K-UNET+SEG, KIKI-Net+SEG, IDSLR+SEG, and KDSLR+SEG are the segmentation performances of the SEG network on the pre-trained reconstruction networks K-UNET, KDSLR, IDSLR, and KIKI-Net, respectively. Among all the methods, the SEG network alone is the best performing since it is trained and tested on fully sampled images. The segmentation performance of other methods is dependent on the reconstruction quality. The IDSLR+SEG provides more accurate labels than other direct cascade methods due to the improved quality of reconstructions (Table \ref{tab:rec_comp}). Our proposed end-to-end training approach, IDSLR-SEG, is seen to offer better performance than the direct cascade methods, including IDSLR+SEG. IDSLR-SEG is trained with labels for all training datasets. The quantitative results in Table \ref{tab:comp_seg} agree with the results in Fig. \ref{fig:brain_seg}. In particular, the proposed approach offers segmentations that are the closest to SEG. The improved segmentation quality of proposed methods over IDSLR+SEG is evident from Fig \ref{fig:brain_seg}.        

\begin{table}[h!]
	\fontsize{8}{16}
	\selectfont
	\centering
	\renewcommand{\arraystretch}{0.8}
	\begin{tabular}{|c|c|c|}
		\hline
		\multicolumn{3}{|c|}{Comparison of Reconstruction} \\ \hline
		Methods& 6-fold & 8-fold \\ \hline
		IDSLR & 21.99 & 20.53  \\
		KDSLR & 21.04 & 20.01 \\
		KIKI-Net & 21.12 & 19.93  \\
		K-UNET & 19.93 & 19.08 \\
		\textbf{Proposed} & 22.76 & 21.32 \\
		\hline
		\end{tabular}
	\caption{Quantitative comparison of reconstruction methods shown in Fig. \ref{fig:brain_rec} with SNR values averaged over 15 subjects in dB.}
	\label{tab:rec_comp} 
\end{table}

\begin{table}[h!]
	\fontsize{8}{16}
	\selectfont
	\centering
	\renewcommand{\arraystretch}{0.8}
	\begin{tabular}{|c|ccc|ccc|}
		\hline
		\multicolumn{7}{|c|}{Comparison of Segmentation} \\ \hline
		Methods & \multicolumn{3}{|c|}{6-fold} & \multicolumn{3}{|c|}{8-fold} \\
		& CSF & GM & WM & CSF &GM &WM \\ \hline
		KDSLR+SEG & 0.714 & 0.749 & 0.823 &0.652 &0.728 &0.801\\
		IDSLR+SEG & 0.738 & 0.751 & 0.841 &0.704  &0.749  &0.803\\ 
		KIKI-Net+SEG & 0.703 & 0.754 & 0.837 &0.637 &0.711 &0.793\\ 
		KUNET+SEG & 0.642& 0.741 &0.788 &0.611 &0.692 &0.747\\
		\textbf{Proposed} &0.793 & 0.841 & 0.898 &0.767 & 0.826 & 0.889\\
		\hline
		& \multicolumn{2}{|c|}{CSF} & \multicolumn{2}{|c|}{GM} & \multicolumn{2}{|c|}{WM} \\ \hline
		SEG & \multicolumn{2}{c}{0.805} & \multicolumn{2}{c}{0.855} & \multicolumn{2}{c|}{0.913} \\ \hline 
		\end{tabular}
	\caption{Comparison of Segmentation Performance. Dice coefficient is reported for CSF, GM and WM segmentation. All the metrics are averaged over 15 subjects. The proposed methods are in bold text.}
	\label{tab:comp_seg} 
\end{table}
\subsection{Benefit of Shared Encoder Architecture}
We study the performance of different joint reconstruction-segmentation architectures in Table \ref{tab:comp_joint}. We note that the difference in performance between the proposed fully supervised and the few-shot learning approaches is marginal, which is also seen from Fig.\ref{fig:brain_seg}.(e). On the other hand, for cascade architecture, the segmentation performance degrades significantly in the few-shot learning case. Note that in few-shot approaches, segmentation labels are available for only 10\% of the data (4 subjects). We also note that this trend is mirrored by the reconstruction performances in Table \ref{tab:comp_joint}. The proposed few-shot approach gives a similar performance compared to the fully supervised one in terms of average SNR.  We note from the zoomed-in sections of Fig. \ref{fig:brain_seg} that the cascaded network misses some key regions of the gray matter at the center of the slice, denoted by blue arrows. The proposed fully supervised and few-shot learning approaches have preserved those segmentation details and perform at par with the SEG network (segmentation on fully sampled images).    
\begin{figure}
	\centering
	\includegraphics[width=0.48\textwidth,keepaspectratio=true,trim={1.8cm 6.8cm 11.3cm 7.2cm},clip]{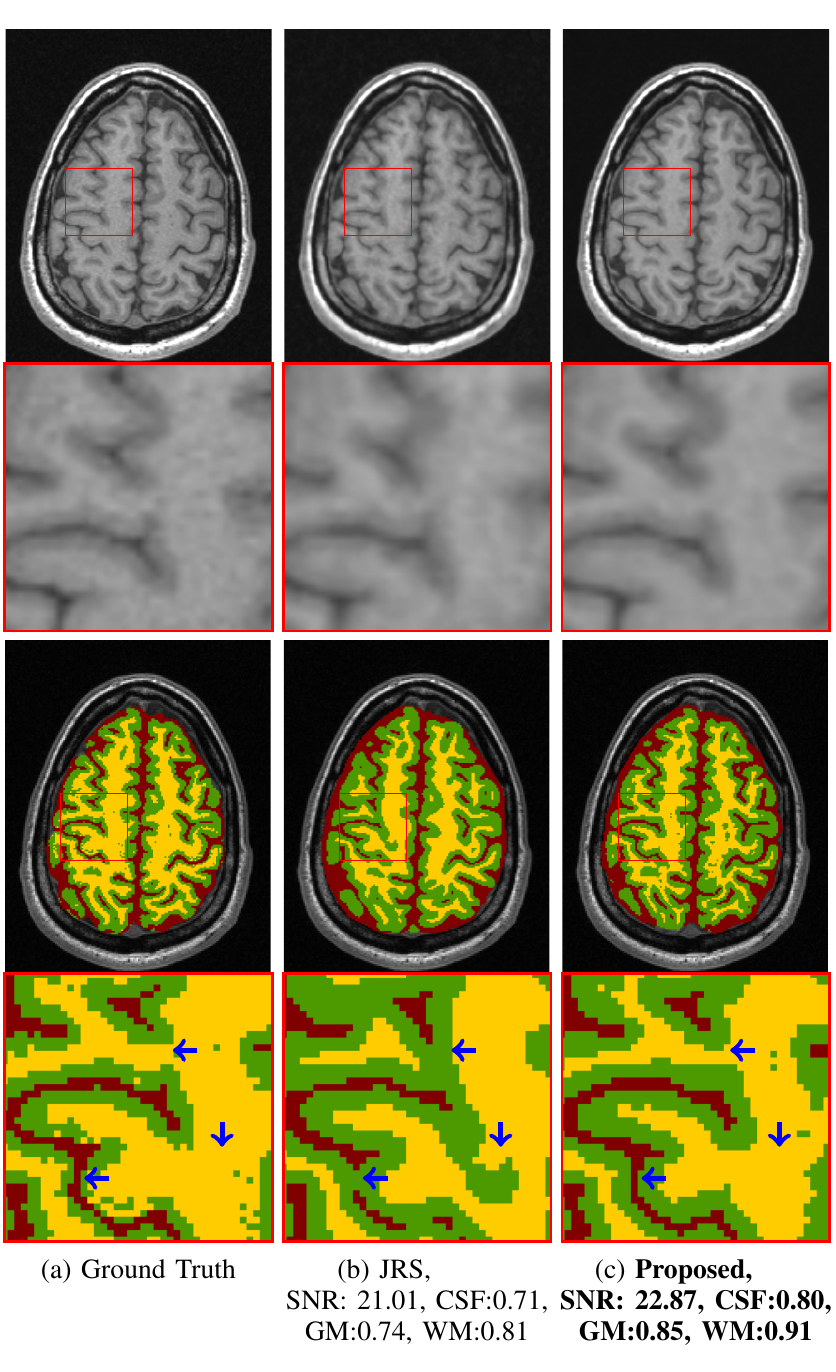}
	\caption{Comparison of JRS \cite{sun2019joint} with the proposed scheme on 6x accelerated multi-channel brain data. SNR (dB) and Dice coefficient values are reported for the particular slice in each case. The data was undersampled using a Cartesian 2D non-uniform variable-density mask. The top row shows reconstructions (magnitude images) and their zoomed-in portions, while the bottom row shows corresponding zoomed-in segmentations. The proposed method outperforms JRS architecture; the key differences are highlighted by zoomed-in images in red boxes.}
	\label{fig:brain_sota}
\end{figure}
  
\begin{table}[h!]
	\fontsize{8}{16}
	\selectfont
	\centering
	\renewcommand{\arraystretch}{0.8}
	\begin{tabular}{|c|c|ccc|}
		\hline
		\multicolumn{5}{|c|}{Comparison of Joint Recon-Seg Architectures} \\ \hline
		\multicolumn{5}{|c|}{6-fold undersampling} \\ \hline
		Methods& SNR & CSF & GM & WM \\ \hline
		\textbf{Proposed} &22.76 & 0.793 & 0.841 & 0.898\\
		\textbf{Proposed (few shot)} &22.71 &0.780 &0.828 &0.895 \\ 
		Cascade & 22.75 &0.795 &0.838 &0.891\\ 
		Cascade (few shot) &22.39 &0.726 &0.773 &0.840 \\ 
		JRS  & 20.89 &0.705 &0.721 &0.796 \\\hline	
		\multicolumn{5}{|c|}{8-fold undersampling} \\ \hline
		\textbf{Proposed} &21.32 & 0.767 & 0.826 & 0.889\\
		\textbf{Proposed (few shot)} &21.28 &0.761 &0.817 &0.886 \\
		Cascade & 21.35 &0.761 &0.818 &0.893\\
		Cascade (few shot) &20.97 &0.711 &0.747 &0.824 \\ 
		JRS  &19.74  &0.683 &0.702 &0.776\\ \hline
		\end{tabular}
	\caption{Comparison of joint architectures for reconstruction and segmentation quality evaluation. Dice coefficient is reported for CSF, GM, and WM segmentation, and the reconstruction SNR values are in dB. All the metrics are averaged over 15 subjects. }
	\label{tab:comp_joint} 
\end{table}
\subsection{Comparison with State-of-the-art}
The proposed framework is compared against the JRS framework in Table \ref{tab:comp_joint} and Fig. \ref{fig:brain_sota}. The JRS was introduced for single-channel MRI; we compare the proposed method against a multi-channel extension of it as shown in Fig. \ref{fig:arch_comp}.(b). The weights of the reconstruction network are not shared across the iterations, while the weights of the segmentation decoder are shared. Hence, if $K$ denotes the number of iterations of the reconstruction network, then there are $K$ times more {\rm train}able parameters than the proposed method. In the JRS scheme, the segmentation output of decoders from all the iterations are linearly combined to obtain a final segmentation. We observe that both the reconstruction and segmentation quality of the proposed scheme are superior to JRS; we note a reduction in blurring and improved preservation of details in the proposed reconstruction. JRS segmentation shows errors in the regions indicated by arrows which are not present in the proposed segmentation. Thus, there is a significant improvement in segmentation quality.

\section{Discussion}
The comparison of the proposed spatial domain deep-SLR scheme (IDSLR) against other calibrationless methods in Table \ref{tab:rec_comp} and Fig. \ref{fig:brain_rec} shows the benefit of the proposed framework, which combines the image domain annihilation relations (IDSLR) with the segmentation task.  The improved performance of KDSLR over K-UNET can be attributed to the model-based strategy that offers improved data consistency. We observe that the proposed IDSLR scheme offers more than 1 dB of improvement over all of the other methods, which can be evidenced by the lower alias artifacts and sharper edges shown in Fig. \ref{fig:brain_rec}. The improved performance over KDSLR \cite{pramanik2020deep} can be attributed to the use of the spatial domain annihilation relations that result from the multi-channel acquisition scheme, as well as those resulting from the redundancy of patches. The additional improvement in the reconstruction performance offered by the proposed scheme, which uses a linear combination of segmentation and reconstruction losses, may be slightly counterintuitive from an optimization perspective. However, this improved performance confirms our hypothesis that the exploitation of the synergies between the segmentation and reconstruction tasks will aid both tasks.  In particular, the addition of the segmentation task encourages the network to learn improved latent representations, resulting in improved preservation of edges and details in the reconstructed images.

The segmentation results in Table \ref{tab:comp_seg} and Fig. \ref{fig:brain_seg} clearly show the benefit of the joint training strategy IDSLR-SEG compared to straightforward cascading of a pre-trained segmentation network with reconstruction algorithms, including IDSLR+SEG. In particular, the SEG scheme that was pre-trained on fully sampled images will offer lower performance in the presence of blurring and alias artifacts. By contrast, the proposed joint training strategy IDSLR-SEG offers improved segmentation, evidenced visually and by the lower Dice scores. The slight degradation in the segmentation accuracy of IDSLR-SEG over the SEG approach, where fully sampled images are used, is the price to pay for the acceleration.

The comparison of the few-shot training setting with only 10\%  of segmentation labels against the previous setting with all the segmentation labels in Table \ref{tab:comp_joint} shows that the proposed architecture with shared encoders offers minimal degradation in segmentation and reconstruction performance.  By contrast, the performance of the cascade approach is seen to degrade with a reduced number of segmentation labels.
The shared encoder enables the segmentation network to reduce generalization errors, even though it is trained with few subjects. Since the encoder is trained with a supervised reconstruction loss from several subjects, it learns a very robust feature representation, thus reducing the risk of overfitting in the segmentation network. In contrast to the few-shot learning of the cascade network, the segmentation algorithm fails to learn enough features due to insufficient training data. We also note that this trend is mirrored by the reconstruction performances in Table \ref{tab:comp_joint}. The proposed few-shot approach gives similar performance to the fully supervised one in terms of average SNR.

\section{Conclusion}
We introduced a novel image-domain deep structured low-rank algorithm (IDSLR) for calibrationless PMRI recovery. The proposed IDSLR scheme is a DL-based non-linear extension of locally low-rank approaches such as CLEAR. The network learns annihilation relations in the spatial domain obtained from the smoothness assumption on coil sensitivities. IDSLR exploits more annihilation relations than the k-space approach KDSLR, thus offering improved performance. To reduce the impact of undersampling on downstream tasks such as MRI segmentation, we proposed a joint reconstruction-segmentation framework trained in an end-to-end fashion. An IDSLR network is combined with a UNET for segmentation such that the encoder is shared between the tasks. The proposed joint training strategy outperforms the direct cascade of image reconstruction and pre-trained segmentation algorithms, in addition to improving the reconstruction performance. The proposed IDSLR-SEG network with a shared encoder for segmentation and reconstruction tasks enables a few-shot learning strategy, which reduces the number of segmented datasets for training by a factor of ten. 
\bibliographystyle{IEEEtran}
\bibliography{refs}
\end{document}